 \documentclass[preprint2]{aastex}

\newcommand{\target}{4U\,0614+091\,}
\newcommand{\porb}{P$_{\rm orb}$}

\shorttitle{Time-resolved photometry of \target }
\shortauthors{Shahbaz et al.}

\begin{document}


\title{Time-resolved optical photometry of the ultra-compact binary \target}


\author{T. Shahbaz\altaffilmark{1}}
\affil{Instituto de Astrofisica de Canarias, 
c/ via Lactea s/n, La Laguna, E38200, Tenerife, Spain}
\email{tsh@iac.es}

\author{C. A. Watson\altaffilmark{2}}
\affil{Department of Physics and Astronomy, University of Sheffield,
Hicks Building, Sheffield, S3 7RH, England}
\email{c.watson@sheffield.ac.uk} 

\author{C. Zurita\altaffilmark{3}}
\affil{Instituto de Astronomia, Universidad Nacional Autonoma de Mexico, 
Ensenada, Mexico}
\email{czurita@astrosen.unam.mx}

\author{E. Villaver\altaffilmark{4}}
\affil{Space Telescope Science Institute, 3700 San Martin Drive, 
Baltimore, MD 21218, USA}
\email{villaver@stsci.edu}

\author{H. Hernandez-Peralta\altaffilmark{1}}
\affil{Instituto de Astrofisica de Canarias, 
c/ via Lactea s/n, La Laguna, E38200, Tenerife, Spain}
\email{hhp@iac.es}


\altaffiltext{4}{Affiliated with the Hubble Space 
Telescope Division of the European Space Agency}


\begin{abstract}

We present a detailed optical study of the ultra-compact X-ray binary
\target. We have used 63\,hrs of time-resolved optical photometry taken with
three  different telescopes (IAC80, NOT and SPM) to search for optical
modulations. 
The power spectra of each
dataset reveals sinusoidal modulations with
different periods, which are not always present.  The strongest modulation has
a period of 51.3\,mins, a semi-amplitude of 4.6\,mmags, and is present in  the
IAC80 data. The SPM and NOT data show periods of 42\,mins and 64\,mins
respectively, but with much weaker amplitudes, 2.6\,mags and 1.3\,mmags
respectively. These modulations arise from either X-ray
irradiation of the inner face of the secondary star and/or a superhump
modulation from the accretion disc, or quasi-periodic modulations in the
accretion disc.  It is unclear whether these  periods/quasi-periodic
modulations are related to the orbital period,  however, the strongest period
of 51.3\,mins is close to earlier tentative orbital periods.
Further observations taken over a long base-line are
encouraged.

\end{abstract}


\keywords{Stars}



\section{Introduction}

Low-mass X-ray binaries (LMXBs) are systems in which a low-mass companion star
transfers material onto a neutron star or black hole. Most of the systems have
orbital periods of between a few hours to days and contain ordinary
hydrogen-rich donor stars. While these systems have  minimum orbital periods
around 80 mins, systems with hydrogen-poor or  degenerate donor stars can,
however,  systems with hydrogen-poor or degenerate donor stars can evolve to
extremely small binary separations with orbital periods  as short as a few
minutes \citep{Nelson86}. Such systems are called ultra-compact X-ray binaries
(UCXBs) and have a range in orbital periods from 11 to 50 minutes (see
\citealt{NJ06})

Finding UCXBs is difficult because measuring orbital periods (\porb) is
generally difficult in LMXBs.  The current known sample consists of 27 systems,
8 with known periods, 4 with tentative periods and 15 candidate systems
\citep{Zand07}. The identification of UCXBs is mostly done through measuring
\porb\ via (1) timing of Doppler-delayed pulses if the accretor is a pulsar;
(2)  measuring  periodic X-ray eclipses/dips if the binary inclination is high
enough and (3) through the measurement of periodic modulations resulting  from
X-ray heating of the inner side of the donor star or from the superhump
phenomenon that is predicted for extreme mass ratio systems. There are also two
indirect methods to identify UCXBs without measuring \porb, which depend on the
fact that in an UCXB the accretion disk are relatively small; (1) for the same
X-ray flux, $M_V$ is about 4 mags fainter for UCXBs than for normal LMXBs and
(2) a method which depends on the  critical accretion rate below which a system
becomes transient; if the persistent accretion luminosity is 1\% of Eddington 
\citep{Zand07}

\target\ is a low luminosity X-ray binary. Thermonuclear type I X-ray bursts
were observed by {\it OSO-8} \citep{Swank78} and the the compact object is a
neutron star. Historically, \target\ has been associated with the brightest
X-ray bursts.  Based on the detection of a bright burst with {\it Watch}, and
Eddington limit  arguments, \citet{Brandt92} argue that the distance to
\target\ is probably  $<$3\,kpc. Several lines of evidence point to the
conclusion that \target is an UCXB. Based on the comparison of the enhanced
neon to oxygen ratio to known  UCXBs, \citet{Juett01} argue that \target\ is
also a UCXB.  Further support is provided by  optical spectroscopy, which has
revealed carbon and oxygen emission lines, but no evidence for hydrogen or
helium  \citep{Nelemans06}. The optical counterpart to \target, V1055\,Ori, is
also intrinsically faint ($V$=18.5). The faintness of its persistent X-ray emission and
nearby distance suggest a low accretion rate, which is consistent with an
orbital period $<$1\,hr \citep{Deloye03}.  In this paper we present the results
on a long term campaign to find a stable period in \target, which would most
likely represent the orbital period.

\section{Observations and Data Reduction}
\label{OBSERVATION}

Our optical photometric observations were taken in 2006 and 2007 using  the
80\,cm IAC80 telescope (Izana, Spain), the 2.5m Nordic Optical Telescope  (NOT;
La Palma, Spain)  and the 1.5 telescope at San Pedro Martin (SPM; Mexico).   A
log of the observations is given in Table\,\ref{Log}.  V-band images were taken
using exposure times ranging   from 40\,sec to 200\,sec depending on the
seeing, weather conditions and the aperture of the telescope being used. Bias
images and Dome flat fields were also  taken for calibration purposes. Standard
stars could not be taken due to non-photometric weather conditions.

We used IRAF for our data reduction, which included bias subtraction using bias
images or the overscan regions of the CCD, and flat-fielding using sky flat
fields taken during twilight. The ULTRACAM reduction  pipeline software
\citep{Dhillon01}  was then used to obtain lightcurves for \target\ and several
comparison stars by extracting the counts using aperture photometry. A variable
aperture which scaled with the seeing was used.  Differential lightcurves were
then obtained by computing the  count ratio of \target\ with respect to a local
standard (the same  non-variable star for each dataset). As a check of the
photometry and systematics in the reduction procedure, we also extracted
lightcurves of a comparison star  similar in brightness to the target.  The
photometric accuracy of \target\ for each exposure is about 2 percent and
agrees with the scatter of the comparison star with similar brightness.

\section{Results}
\label{RESULTS}

Figure\,1 shows the lightcurve for \target\ obtained from each observing
site.   The data clearly  exhibits a periodic modulation, which changes
strength with time. To search for periodic/quasi-periodic modulations we use 
the method of  Lomb-Scargle \citep{Press92} to compute the  periodograms of all
the datasets, using the constraints imposed by the Nyquist frequency  and the
typical duration of each observation. The results are shown in
Figure\,2. We also show the 99 percent confidence level, which allows us to
demonstrate the significance of the peaks detected in the Lomb-Scargle
periodogram. The level was calculated from a Monte Carlo simulation, which
using 10,000 sets of Gaussian noise with mean and variance taken from 
the dataset. The periodogram for each datasets shows many peaks significant at
the 99 percent level. The strongest peak in the 
the IAC80 data is at 51.3\,mins,
whereas the NOT data shows a weaker peak at 64.1\,mins, and the SPM data shows
an even  weaker  modulation at 42\,mins.  
The semi-amplitudes of the modulation
is 4.6\,mmags, 2,6\,mmags and  1.3\,mmags for the IAC80, SPM and NOT data
respectively. We have looked at the RXTE ASM lightcurves and the X-ray flux
seems to be  similar on the dates of our optical observations. 

\section{Discussion}
\label{DISCUSSION}

\citet{Nelemans04} obtained VLT optical spectra of \target\ and identified
features of  relatively low ionisation states of carbon and oxygen. This
clearly identifies \target\ as an UCXB and suggests that the donor in the
system in a carbon-oxygen white dwarf. For \target\ and 4U\,1626-67 there are
clear indications that the discs are dominated by C and O. \citet{Werner06}
have also obtained VLT spectra and compared them with detailed NLTE models for
spectra of UCXBs. Unfortunately, the NTLE models do not sufficiently agree with
the observed spectra  for quantitative abundance analysis. Although simple LTE
models seem to fit the data better, they also cannot be used for quantitative
measurements because NLTE effects mainly due to X-ray irradiation, need to be
taken into account. 

We have found  several sinusoidal modulation  in the  optical lightcurve of
\target. These modulations most likely arise from either X-ray irradiation of
the inner face of the secondary star and/or a superhump modulation from the
accretion disc, or quasi-periodic oscillations in the accretion disc.   This is
not surprising as UCXBs are known to show orbital modulations as well as QPOs.
e.g. strong 15 minute optical/UV quasi-periodic oscillations were previously
detected in the 42\,min UCXB 4U\,1626--67 \citep{Chak01}, showing that
photometric variability in an UCXB need not only occur near the orbital
period.  

\citet{OBrien05} have reported a 50\,min periodicity in their high-speed
optical data taken with ULTRACAM.   Time-resolved VLT spectroscopy also shows
some evidience, although marginal, for a 49\,min periodicity in the  weak
absorption line near 4960\AA\ \citep{Nelemans06}. The strongest period we detect
is at 51.3\,mins and is present in the IAC80  data. This may reflect the
superhump period, slightly longer than the orbital period. However,  only
observations long enough to contain many modulation cycles can distinguish
between a periodic and a quasi-periodic modulation and allow a secure
measurement of the orbital period.

\acknowledgments

TS acknowledges support from the Spanish Ministry of Science
under the programme Ram\'{o}n y Cajal.
TS and MD acknowledges support from the Spanish Ministry of Science  and
Technology  under the programmes AYA\,2004\,02646 and AYA\,2007\,66887.
CAW is supported by a PPARC postdoctoral fellowship.



{\it Facilities:} \facility{IAC80}, \facility{NOT}, \facility{SPM}.

\clearpage

\begin{table*}
\begin{center}
\caption{Log of observations.}
\label{Log}
\begin{tabular}{lcccc}
\tableline\tableline
Date     & Telescope & exp. time   & duration    & avg. seeing \\ 
\tableline
10/11/06 &  IAC80    &  100\,sec   &  316\,mins  &  1.6\arcsec \\
13/12/06 &  IAC80    &  200\,sec   &  410\,mins  &  1.9\arcsec \\
14/12/06 &  IAC80    &  200\,sec   &  334\,mins  &  1.7\arcsec \\
15/12/06 &  IAC80    &  200\,sec   &  354\,mins  &  1.5\arcsec \\ 
\tableline 
 8/1/07  &  SPM      &  100\,sec   &  205\,mins  &  4.1\arcsec \\
 9/1/07  &  SPM      &  100\,sec   &  411\,mins  &  4.1\arcsec \\
11/1/07  &  SPM      &  100\,sec   &  391\,mins  &  5.4\arcsec \\ 
\tableline
12/2/06  &  NOT      &   40\,sec   &  296\,mins  &  1.7\arcsec \\
13/2/06  &  NOT      &   40\,sec   &  238\,mins  &  1.5\arcsec \\
14/2/06  &  NOT      &   40\,sec   &  271\,mins  &  1.4\arcsec \\
15/2/06  &  NOT      &   40\,sec   &  271\,mins  &  2.4\arcsec \\
16/2/06  &  NOT      &   40\,sec   &  258\,mins  &  1.7\arcsec \\ 
\tableline
\end{tabular}
\end{center}
\end{table*}

\clearpage

\begin{figure*}
\label{fig1}
\begin{center}
\includegraphics[height=5.4cm,angle=-90]{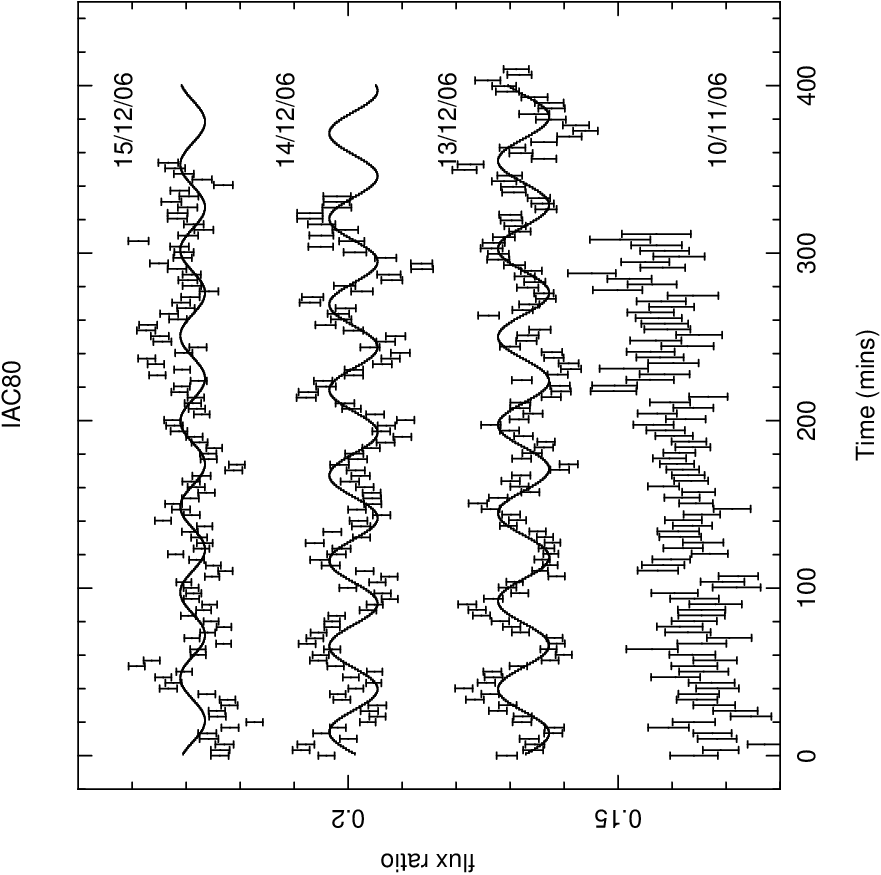}
\includegraphics[height=5.4cm,angle=-90]{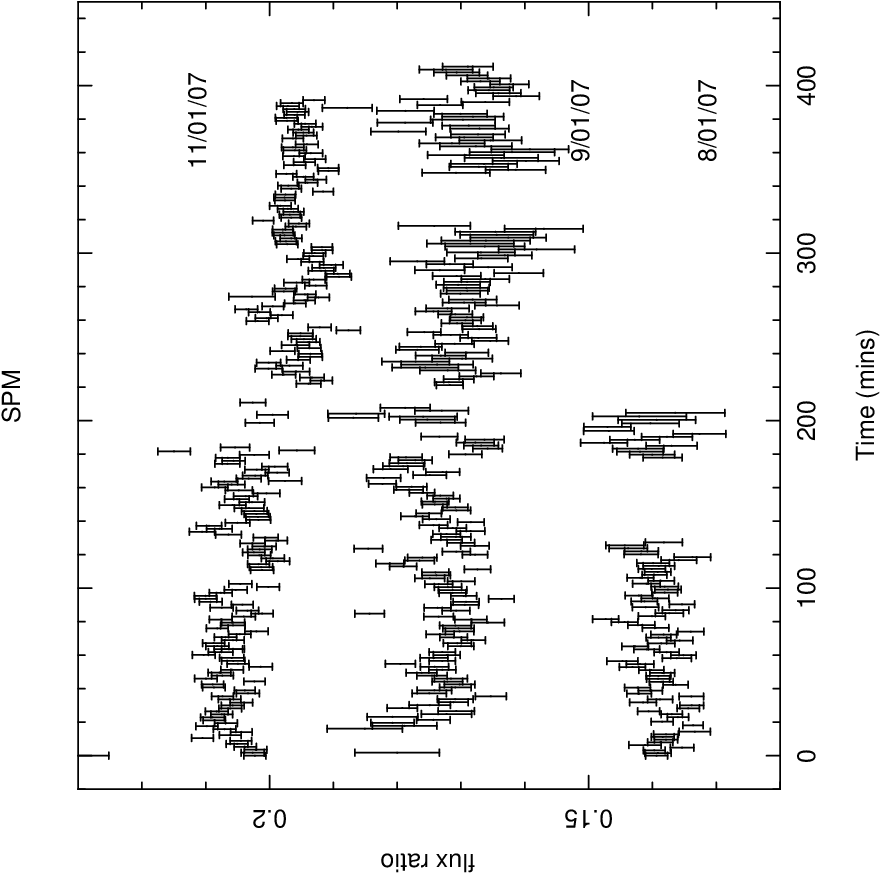}
\includegraphics[height=5.4cm,angle=-90]{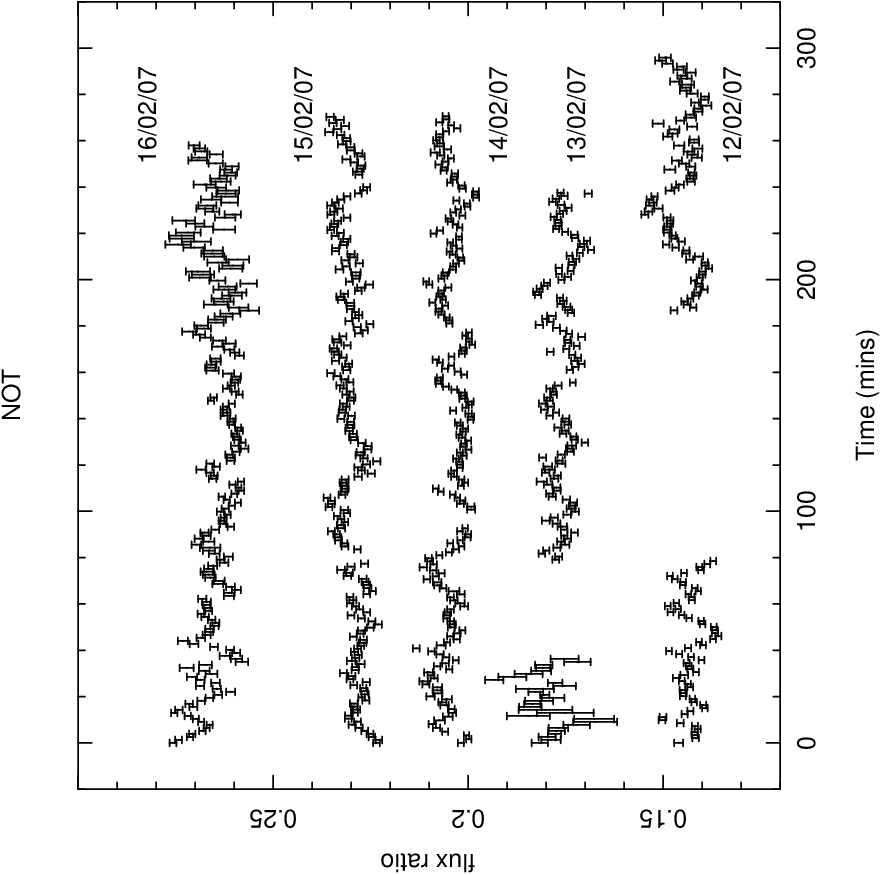}
\caption{V-band light curves of \target\ taken with the 80cm IAC80 telescope 
(left), 1.5m telescope at San Pedro Martin (middle) and the 2.5m NOT (right).
For the IAC80 data we also show a 51.3\,min sinusoidal modulation (solid
curve).}
\end{center}
\end{figure*}

\begin{figure*}
\label{fig2}
\begin{center}
\includegraphics[height=12.5cm,angle=-90]{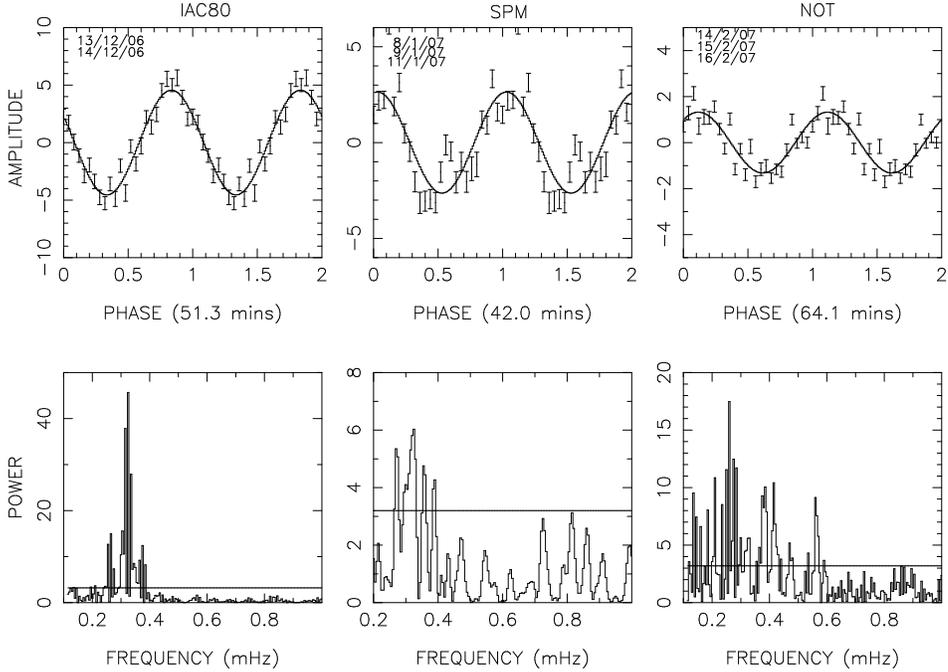}
\caption{The power spectrum (bottom panels) of the lightcurve taken with the 
IAC80, NOT and SPM telescopes. The dates in each panel show the data that  were
used to compute the power spectrum. A Monte Carlo simulations provides the
99 percent confidence level  indicated by the solid horizontal line.
The top panels show the lightcurve folded
and binned on the significant period found for the data from each telescope.}
\end{center}
\end{figure*}

\end{document}